\documentclass[twocolumns]{aastex631}
\usepackage{amssymb}

\begin{document}
\def\func#1{\mathop{\rm #1}\nolimits}
\def\unit#1{\mathord{\thinspace\rm #1}}

\title{iPTF14hls in the circumstellar medium interaction model: A promising
candidate for a pulsational pair-instability supernova}
\correspondingauthor{Ling-Jun Wang, Liang-Duan Liu, Wei-Li Lin}
\email{wanglingjun@ihep.ac.cn, liuld@ccnu.edu.cn, linwl@mail.tsinghua.edu.cn}

\author[0000-0002-8352-1359]{Ling-Jun Wang}
\affiliation{Astroparticle Physics, Institute of High Energy Physics,
Chinese Academy of Sciences, Beijing 100049, China}

\author[0000-0002-8708-0597]{Liang-Duan Liu}
\affiliation{Institute of Astrophysics, Central China Normal University, Wuhan 430079, China;}
\affiliation{Key Laboratory of Quark and Lepton Physics (Central China Normal University), Ministry of Education, Wuhan 430079, China}

\author{Wei-Li Lin}
\affiliation{Physics Department and Tsinghua Center for Astrophysics,
 Tsinghua University, Beijing 100084, China}

\author[0000-0002-7334-2357]{Xiao-Feng Wang}
\affiliation{Physics Department and Tsinghua Center for Astrophysics,
 Tsinghua University, Beijing 100084, China}
\affiliation{Beijing Planetarium, Beijing Academy of Sciences and
Technology, Beijing, 100044, China}

\author[0000-0002-7835-8585]{Zi-Gao Dai}
\affiliation{CAS Key Laboratory for Research in Galaxies and Cosmology, Department of Astronomy, University of Science and Technology of China, Hefei 230026, 
China}

\author{Bing Li}
\affiliation{Astroparticle Physics, Institute of High Energy Physics,
Chinese Academy of Sciences, Beijing 100049, China}

\author{Li-Ming Song}
\affiliation{Astroparticle Physics, Institute of High Energy Physics,
Chinese Academy of Sciences, Beijing 100049, China}

\begin{abstract}
iPTF14hls is a luminous Type II supernova (SN) with a bumpy light curve that
remains debated for its origin. It maintains roughly a constant effective
temperature and luminosity since discovery for about 600 days, followed by a
slow decay. On $\sim 1000$\ days post discovery the light curve transitions
to a very steep decline. A spectrum taken during this steep decline phase
shows clear signatures of shock interaction with dense circumstellar medium
(CSM). Here we explore the possibility of iPTF14hls as an
interaction-powered SN. The light curve of iPTF14hls can be fitted with
wind-like CSMs. Analytic modeling indicates that iPTF14hls may have
undertaken six episodes of mass loss during the last $\sim 200\unit{yr}$.
Assuming that the 1954 eruption triggered the last mass-loss episode, the
stellar-wind velocity is determined to be $40-70\unit{km}\unit{s}^{-1}$,
depending on different models. Mass loss rates are in the range $%
0.4-3.3M_{\odot }\unit{yr}^{-1}$. The inferred total mass of ejecta and CSMs
($M_{\mathrm{ej}}+M_{\mathrm{CSMs}}\simeq 245M_{\odot }$) supports the idea
that iPTF14hls may be a candidate for a (pulsational) pair-instability SN.
Discovery and observations of more similar stellar explosions will help
understand these peculiar SNe.
\end{abstract}

\keywords{stars: massive --- supernovae: general --- supernovae: individual
(iPTF14hls)}

\section{Introduction}

\label{sec:Intro}

iPTF14hls is a nearby ($z=0.0344$) peculiar supernova (SN) first discovered
in $R$ band on 2014 Sep. 22 UT by the Intermediate Palomar Transient Factory
(iPTF) wide-field camera survey \citep{Arcavi17}. It was classified as a
Type II-P \citep{Filippenko97} SN according to its spectroscopic features %
\citep{Li15}. The light curves of iPTF14hls last for more than 1200 days %
\citep{Sollerman19} and are bumpy in the first 500 days since discovery. The
velocities of hydrogen lines of iPTF14hls decline from $8000\unit{km}\unit{s}%
^{-1}$ to $6000\unit{km}\unit{s}^{-1}$ over 600 days, during which the iron
lines (Fe \textsc{ii }$\lambda 5169$) stay at a constant velocity of $4000%
\unit{km}\unit{s}^{-1}$. The effective temperature of iPTF14hls is roughly
constant ($T\approx 6000\unit{K}$). Spectral evolution of iPTF14hls is also
very slow compared to typical SNe II-P \citep{Arcavi17}.

After 600 days from discovery, the evolution of iPTF14hls speeds up. In
particular, the light curve transitions to a very steep decline since $\sim$%
1000 days after discovery \citep{Sollerman19}. During this time a spectrum
taken on day 1153 after discovery reveals a clear signature of shock
interaction with dense circumstellar medium \citep[CSM;][]{Andrews18}.
Interestingly, \cite{Yuan18} report the detection of a variable $\gamma $%
-ray source in the energy band $0.2$ to $500\unit{GeV}$, which is
positionally and temporally consistent with iPTF14hls. However, it cannot be
confirmed whether this $\gamma $-ray source is associated with iPTF14hls or
a blazar candidate.

Several models have been proposed or discussed to interpret iPTF14hls 
\citep{Andrews18, Chugai18, Dessart18, Soker18, Wang18, Woosley18, Gofman19,
Liu19, Modjaz19, Kaplan20, Moriya20, Uno20}. However, till now no single
model is able to interpret all the observational features of iPTF14hls. The
late-time spectrum of iPTF14hls exhibits a double-peak intermediate-width H$%
\alpha $ line indicating strong ejecta-CSM interaction (CSI). Together with
the early bumps and the late steep decline of the light curves of iPTF14hls, 
\cite{Andrews18} proposed that variations in density structure of the CSM
could explain the multiple peaks of the light curve. But they did not model
the long-lived light curve of iPTF14 based on successive collisions of
circumstellar shells and SN ejecta in detail. In previous studies, the CSI
model has been applied to a variety of SNe with bumpy light curves %
\citep{WangLiu2016, Liu18, Li20}. Observations reveal that bumpy features
are very common \citep{Hosseinzadeh21} in the light curves of superluminous
SNe \citep[SLSNe;][]{Quimby11, GalYam12,
GalYam19, Moriya18, Inserra19}. This indicates that mass loss of massive
stars during their final evolution stages are common \citep{Smith14} if the
bumps in the light curves of SLSNe are caused by CSM interactions.
Signatures for the presence of nearby CSMs have also been reported for
several SNe II-P/II-L/IIb 
\citep{Quimby07, GalYam14, Khazov16, Yaron17,
Hosseinzadeh18, Rui19, Bruch21, Jacobson-Galan22} by the detection of
flash-ionized emission lines from dense circumstellar wind.

The CSI model is also able to produce some rapidly varying light curves that
can be applied to the emerging fast blue optical transients 
\citep{Tolstov19, WangSQ19, Leung20, Leung21, Xiang21, GalYam22,
Pellegrino22}. This is also an appealing factor of the CSI model because the
third peak (see Figure \ref{fig:lc-out} below) in the light curve of
iPTF14hls varies so rapidly that it can be fitted only by a CSM interaction
or other instantaneously released energy \citep{Wang18,
Kaplan20}.

Based on above considerations, here we explore the possibility of
interpreting iPTF14hls as a multiple interaction-powered SN. We apply the
semi-analytic multiple CSI model \citep{Liu18} to iPTF14hls to infer its
explosion energy, ejecta mass, and CSM parameters, which allow us to
consider the mass-loss history of the progenitor system. The semi-analytic
model is based on a spherically symmetric explosion with optically thick
CSMs in spheric geometry \citep{Chatzopoulos12}. As a result, the model
predicts a stationary photosphere, inconsistent with the observed
photospheric velocities. However, such a model can be improved by
considering a clumpy CSM \citep{Chugai94} or an asymmetric, disk-like medium
embedded within the spherically expanding ejecta 
\citep{Smith17, Andrews18,
Suzuki19}. The semi-analytic model is computationally efficient and can give
a convenient estimate of the relevant parameters.

This paper is structured as follows. The models and fitting results are
presented in Section \ref{sec:model}. Mass loss history is derived in
Section \ref{sec:mass-loss}. Discussion and conclusions are given in Section %
\ref{sec:conc}.

\section{Models and results}

\label{sec:model}

\subsection{CSI model}

The CSI model assumes that the ejecta expand at a constant velocity $v_{%
\mathrm{SN}}$ with an inner power-law density profile ($\rho \propto
r^{-\delta }$ with $\delta <3$) surrounded by a steep outer power-law
profile 
\citep[$\rho \propto r^{-n}$ with $n>5$;][]{Parker61, Chevalier82,
Chevalier94}. The Lagrangian coordinate for the transition from the inner
flat profile to the outer steep profile is denoted as $x_{0}$ %
\citep{Chatzopoulos12}. In this paper, we set $\delta =0$, $x_{0}=0.7$.

The density profile of CSMs is also assumed in a power-law form $\rho
\propto r^{-s}$. The CSM was ejected by the SN progenitor during its
evolution to the final collapse. Usually two kinds of density profile of
CSMs are considered, namely the stellar wind with $s=2$ or a uniform shell
with $s=0$. In a multiple CSI model \citep{Liu18}, we denote the mass of the 
$i$th CSM as $M_{\mathrm{CSM},i}$. The inner radius of the $i$th CSM is
denoted as $R_{\mathrm{in},i}$. Similarly, the CSM density at $R_{\mathrm{in}%
,i}$ is denoted as $\rho _{\mathrm{in},i}$.

Collision of the ejecta with CSM results in the formation of a forward shock
(FS) propagating into CSM and a reverse shock (RS) propagating into the
ejecta. The dynamical evolution of FS and RS is described by a self-similar
solution \citep{Chevalier82}. Both FS and RS convert the kinetic energy of
the homologously expanding ejecta into radiation that powers the SN light
curve. Based on those self-similar solutions one can derive the expression
for the FS and RS input luminosity that is produced from CSI. The input
luminosities of FS and RS have the same temporal index %
\citep{Chatzopoulos12, MoriyaMaeda13, Wang19} 
\begin{equation}
L_{\mathrm{inp,FS}}(t)=L_{\mathrm{FS},\mathrm{tr}}\left( \frac{t}{t_{\mathrm{%
tr}}}\right) ^{\alpha },\qquad L_{\mathrm{inp,RS}}(t)=L_{\mathrm{RS},\mathrm{%
tr}}\left( \frac{t}{t_{\mathrm{tr}}}\right) ^{\alpha },  \label{eq:L_R-L_F}
\end{equation}%
where the temporal index is 
\begin{equation}
\alpha =\frac{2n+6s-ns-15}{n-s},
\end{equation}%
and $t_{\mathrm{tr}}$\ is the characteristic time of dynamical evolution, $%
L_{\mathrm{inp,FS}}$\ and $L_{\mathrm{inp,RS}}$\ are the characteristic
input luminosities of FS and RS, respectively. These characteristic physical
quantities depend on the nature of the CSM and SN explosion, and the
specific expressions are given in \cite{Liu20}. The heating from FS
terminates when all the available optically thick CSM has been swept up, by
which one can obtain the FS termination timescale $t_{\mathrm{FS,\ast }}$.
Similarly, the RS termination time $t_{\mathrm{RS,\ast }}$\ is determined
when all the ejecta $M_{\mathrm{ej}}$\ has been swept.

The input luminosities of FS and RS from $i$th CSM interaction is%
\begin{eqnarray}
L_{\mathrm{{inp,CSM},}i\mathrm{,FS}}(t) &=&\epsilon _{\mathrm{FS},i}L_{%
\mathrm{inp,FS},i}(t), \\
L_{\mathrm{{inp,CSM},}i\mathrm{,RS}}(t) &=&\epsilon _{\mathrm{RS},i}L_{%
\mathrm{inp,RS},i}(t),
\end{eqnarray}%
respectively. Here $\epsilon _{\mathrm{FS}}$\ and $\epsilon _{\mathrm{FS}}$\
are the respective conversion efficiency from the kinetic energy to
radiation by FS and RS. \cite{Chatzopoulos12} assume that all kinetic energy
of SN ejecta converts efficiently to radiation, that is $\epsilon _{\mathrm{%
FS}}=\epsilon _{\mathrm{RS}}=1$. It may be reasonable for the case $M_{%
\mathrm{CSM}}\gg M_{\mathrm{ej}}$, on the contrary, this assumption is
unrealistic for $M_{\mathrm{CSM}}\ll M_{\mathrm{ej}}$\ 
\citep{vanMarle10,
MoriyaBlinnikov13}. Because of poor knowledge of the physical process of
converting the kinetic energy to radiation, we set $\epsilon _{\mathrm{FS}}$%
\ and $\epsilon _{\mathrm{RS}}$\ as independent parameters. After traversing
a CSM, the ejecta will grow in mass and collide with the next CSM. Thus, the
ejecta mass of the $i$th interaction is 
\begin{equation}
M_{\mathrm{ej},i}=M_{\mathrm{ej},i-1}+M_{\mathrm{CSM},i-1}
\end{equation}

Previous studies of multiple-interaction models \citep{Liu18,Li20} have
usually assumed that a RS develops at every shell collision. However,
whether a RS develops depends on the density structure and the pre-shock
pressure. For the first collision of a \textquotedblleft
cold\textquotedblright\ SN ejecta and a \textquotedblleft
cold\textquotedblright\ CSM, both FS and RS would develop. Due to the
pre-heat high pressure of the first FS, in successive collisions of CSM
shells, RSs do not always develop or they are at least very weak with less
effective contributions to the total luminosity. Therefore, in our
calculation, we only consider the contributions from FSs and the first RS
and neglect the contributions from the rest RSs.

Considering the expansion velocity of the CSM is much lower than the typical
velocity of the SN ejecta, a fixed photosphere inside the CSM is usually
adopted \citep{Chatzopoulos12,Liu18}. Under this assumption, the theoretical
bolometric light curves contributed by FS and RS of the $i$th collision are
given by%
\begin{eqnarray}
L_{i,\mathrm{FS}}(t) &=&\frac{1}{t_{\mathrm{diff},i,\mathrm{FS}}}\exp \left(
-\frac{t}{t_{\mathrm{diff},i,\mathrm{FS}}}\right) \int_{0}^{t}\exp \left( 
\frac{t^{\prime }}{t_{\mathrm{diff},i,\mathrm{FS}}}\right) L_{\mathrm{inp},%
\mathrm{CSM},i,\mathrm{FS}}\left( t^{\prime }\right) dt^{\prime }, \\
L_{i,\mathrm{RS}}(t) &=&\frac{1}{t_{\mathrm{diff},i,\mathrm{RS}}}\exp \left(
-\frac{t}{t_{\mathrm{diff},i,\mathrm{RS}}}\right) \int_{0}^{t}\exp \left( 
\frac{t^{\prime }}{t_{\mathrm{diff},i,\mathrm{RS}}}\right) L_{\mathrm{inp},%
\mathrm{CSM},i,\mathrm{RS}}\left( t^{\prime }\right) dt^{\prime },
\end{eqnarray}%
respectively. Here $t_{\mathrm{{diff},}i\mathrm{,FS}}$\ and $t_{\mathrm{{diff%
},}i\mathrm{,RS}}$\ are the respective diffusion times of the FS and RS of
the $i$th CSM. The theoretical bolometric light curve of $N$\ times
interactions is given by%
\begin{equation}
L\left( t\right) =\sum_{i=1}^{N}\left[ L_{i,\mathrm{FS}}(t)+L_{i,\mathrm{RS}%
}(t)\right] .
\end{equation}%
After $t_{i,\mathrm{FS,\ast }}$\ and $t_{i,\mathrm{RS,\ast }}$, luminosity
input from FS and RS ceases in the CSI models, and the decline in luminosity
is predicted to be exponential.

To handle the diffusion of the shock energy, different authors take
different recipes. \cite{Liu18} adopted diffusion timescale in the optically
thick CSM for both FS and RS, that is%
\begin{equation}
t_{\mathrm{{diff},}i\mathrm{,FS}}=t_{\mathrm{{diff},}i\mathrm{,RS}%
}=\sum_{j=i}^{N}\frac{\kappa _{j}M_{\mathrm{CSM},\mathrm{th},j}}{\beta cR_{%
\mathrm{ph}}},
\end{equation}%
where $N$\ is the total number of CSMs, $\beta \simeq 13.8$\ is a constant %
\citep{Arnett82}, $c$\ is the speed of light, $R_{\mathrm{ph}}$\ is
photospheric radius of the outermost CSM, $\kappa _{j}$, and \ $M_{\mathrm{%
CSM},\mathrm{th},j}\simeq M_{\mathrm{CSM},j}$\ are the optical opacity and
optically thick mass of the $j$th CSM. We denote recipe used in \cite{Liu18}
as CSI-L.

\cite{Wheeler17} assume that the shocks locate in the center of the ejecta,
that is, the original spirit of \cite{Arnett82}. In this situation, the
diffusion time of the $i$th CSI is%
\begin{equation}
t_{\mathrm{{diff},i,FS}}=t_{\mathrm{{diff},i,RS}}=\frac{\kappa M_{\mathrm{ej}%
}}{\beta cR_{\mathrm{ph}}}+\sum_{i=1}^{N}\frac{\kappa _{i}M_{\mathrm{CSM},%
\mathrm{th},i}}{\beta cR_{\mathrm{ph}}},
\end{equation}%
where $\kappa $\ is the optical opacity of the ejecta. To be distinctive,
hereafter we denote the recipe in \cite{Wheeler17} as CSI-W.

However, the FS and RS are not centrally concentrated and move within the
CSM and ejecta, respectively. The assumption of a deeply located power
source is approximately appropriate for the RS, but is not true for the FS.
As one may expect, different recipes have different advantages and
shortcomings. The CSI-W model can better handle the RS luminosity, while the
CSI-L model handles the FS luminosity better. The model can be improved by
combining the advantages of models CSI-L and CSI-W. We can set the diffusion
time of the FS as done in the CSI-L model, while setting the diffusion time
of RS as done in the CSI-W model, that is 
\begin{eqnarray}
t_{\mathrm{{diff},i,FS}} &=&\sum_{j=i}^{N}\frac{\kappa _{j}M_{\mathrm{CSM},%
\mathrm{th},j}}{\beta cR_{\mathrm{ph}}}, \\
t_{\mathrm{{diff},i,RS}} &=&\frac{\kappa M_{\mathrm{ej}}}{\beta cR_{\mathrm{%
ph}}}+\sum_{i=1}^{N}\frac{\kappa _{i}M_{\mathrm{CSM},\mathrm{th},i}}{\beta
cR_{\mathrm{ph}}}.
\end{eqnarray}%
Such a treatment is denoted as model CSI-LW and the Python code is available
on GitHub\footnote{\texttt{CatFit} codebase: %
\url{https://github.com/Lingjun-Wang/CatFit}.} and version 1.0 is archived
in Zenodo \citep{Wang22}.

\begin{figure*}[tbph]
\includegraphics[width=1\textwidth,angle=0,width=15cm]{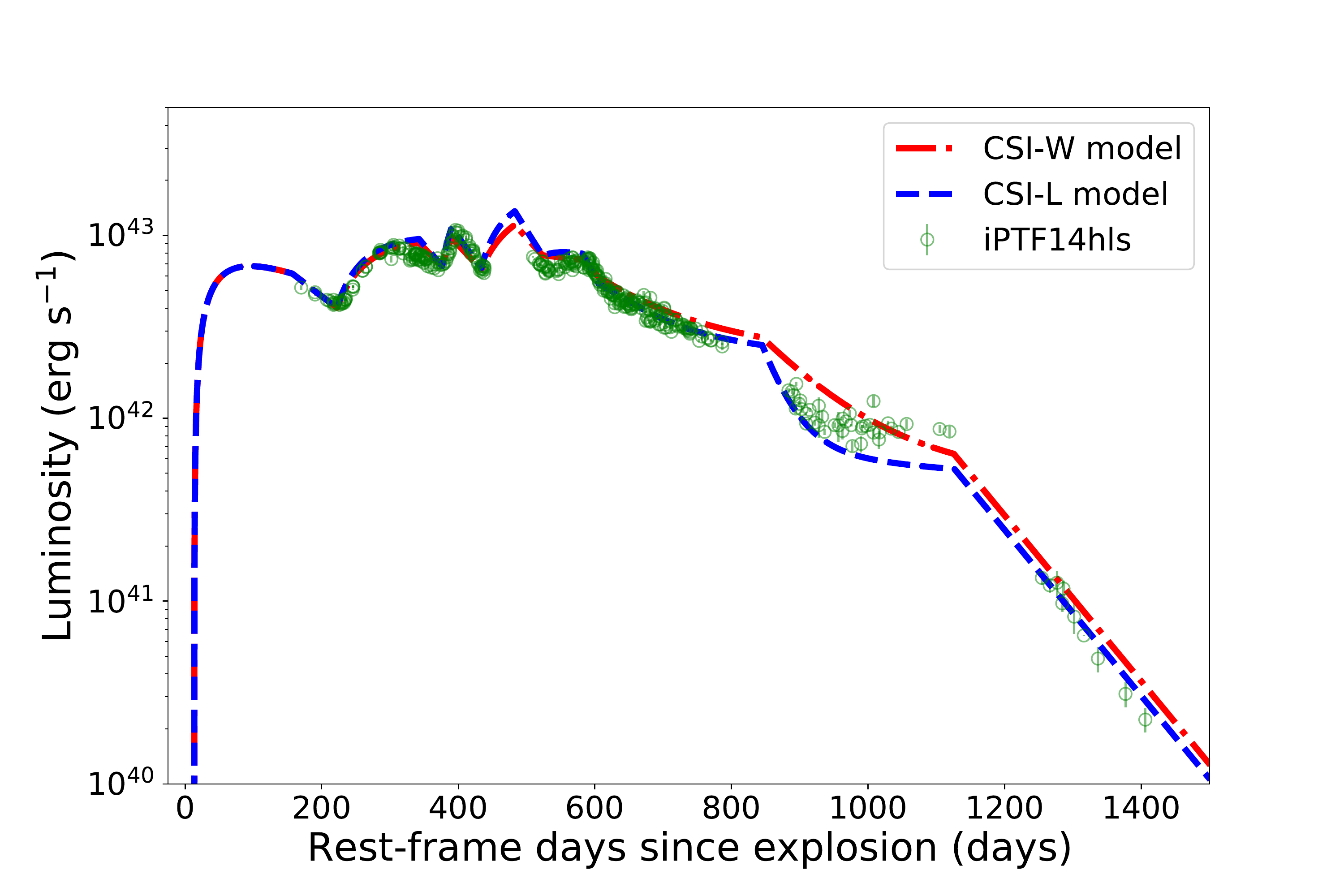} 
\centering
\caption{Fit to the bolometric luminosity of iPTF14hls by the CSI-L model
and CSI-W model. The data are taken from \protect\cite{Arcavi17} and 
\protect\cite{Sollerman19}.}
\label{fig:lc-out}
\end{figure*}

\begin{table*}[tbph]
\caption{Model parameters for iPTF14hls based on the pure-CSI models.}
\label{tbl:para-L-W}
\begin{center}
\begin{tabular}{ccccccc}
\hline\hline
ejecta properties &  &  &  &  &  &  \\ \hline
$M_{\mathrm{ej}} (M_{\odot})$ & 75 &  &  &  &  &  \\ 
$v_{\mathrm{sn}}$ $\left( \unit{km} \unit{s}^{- 1} \right)$ & 5200 &  &  & 
&  &  \\ 
$\delta$ & 0 &  &  &  &  &  \\ 
$n$ & 7 &  &  &  &  &  \\ \hline
CSM properties &  &  &  &  &  &  \\ 
& 1st & 2nd & 3rd & 4th & 5th & 6th \\ \hline
$M_{\mathrm{CSM}} (M_{\odot})$ & 20 & 25.5 & 8 & 25 & 22 & 70 \\ 
$R_{\mathrm{in}} (10^{15} \unit{cm})$ & 0.6 & 10 & 16 & 18 & 21 & 23.5 \\ 
$\rho_{\mathrm{in}} \left( 10^{- 15} \unit{g } \unit{cm}^{- 3} \right)$ & 
1200 & 7.2 & 6.0 & 5.0 & 3.2 & 2.2 \\ 
$\epsilon_{\mathrm{FS}}$ & 0.02 & 0.09 & 0.26 & 0.2 & 0.09 & 0.04 \\ 
$\epsilon_{\mathrm{RS}}$ & 0.4 & 0 & 0 & 0 & 0 & 0 \\ \hline
\end{tabular}%
\end{center}
\end{table*}

The comparison of the fitting results of CSI-L model and CSI-W model is
given in Figure \ref{fig:lc-out}, and the fitting parameters are listed in
the Table \ref{tbl:para-L-W}. With the same fitting parameters, the
theoretical light curve of model CSI-LW is shown in Figure \ref%
{fig:lc-in-out}. The model parameters were varied manually until reasonable
agreement between the observational data and the theoretical model was
found. The initial slow decline of the light curve favors a small $n<9$ and
wind-like CSMs. So we set $n=7$ and $s=2$. With these values, $\alpha =-0.6$%
. The spike (the third peak) in the light curve at $\sim 400$ days after
explosion can only be fitted by a CSM much lighter than the ejecta. The
opacity of the ejecta and CSMs is $\kappa =0.34\unit{cm}^{2}\unit{g}^{-1}$ %
\citep{Matzner99}, which is the Thomson electron scattering opacity for
fully ionized solar metallicity material. We find that in order to get a
close fit to the light curve of iPTF14hls, six wind-like CSM-shells are
required. The CSI model predicts an exponential decay of the light curve
after the forward shock sweeps up all the CSM or the reverse shock sweeps up
all the ejecta. The exponential decay of the light curve starting from $\sim
1100$\ days after explosion is contributed by the first RS (see Figure \ref%
{fig:lc-out}).

Given the same fitting parameters, Figure \ref{fig:lc-out} shows that the
theoretical light curve of CSI-L is brighter than that of CSI-W during the
early phase of the light curve and becomes dimmer during late stage. This
can be easily understood because the diffusion time in CSI-L is shorter and
emission diffuses out earlier.

One of the most influential parameters on the theoretical light curve is the
inner radius of the $i$th CSM $R_{\mathrm{{in},}i}$, which determines the
onset time of the interactions. In CSI-W, ejecta mass $M_{\mathrm{ej}}$\
affects the maximum and the width of the light curve, higher ejecta mass
leading to long-lived RS emission. The larger the CSM mass, the longer the
light curve. For pure CSI model the first RS has a long duration and
contributes mainly to the first peak and the late-time steep decline. The
explosion energy of iPTF14hls is estimated to be 
\citep{Chevalier94,
Chatzopoulos12} 
\begin{equation}
E_{\mathrm{SN}}=\frac{\left( 3-\delta \right) \left( n-3\right) }{2\left(
5-\delta \right) \left( n-5\right) }M_{\mathrm{ej}}\left( x_{0}v_{\mathrm{SN}%
}\right) ^{2}=1.2\times 10^{52}\unit{erg},  \label{eq:exp-energy}
\end{equation}%
of which $5.5\times 10^{50}\unit{erg}$ was emitted as visual light.

\begin{figure*}[tbph]
\includegraphics[width=1\textwidth,angle=0,width=8.9cm]{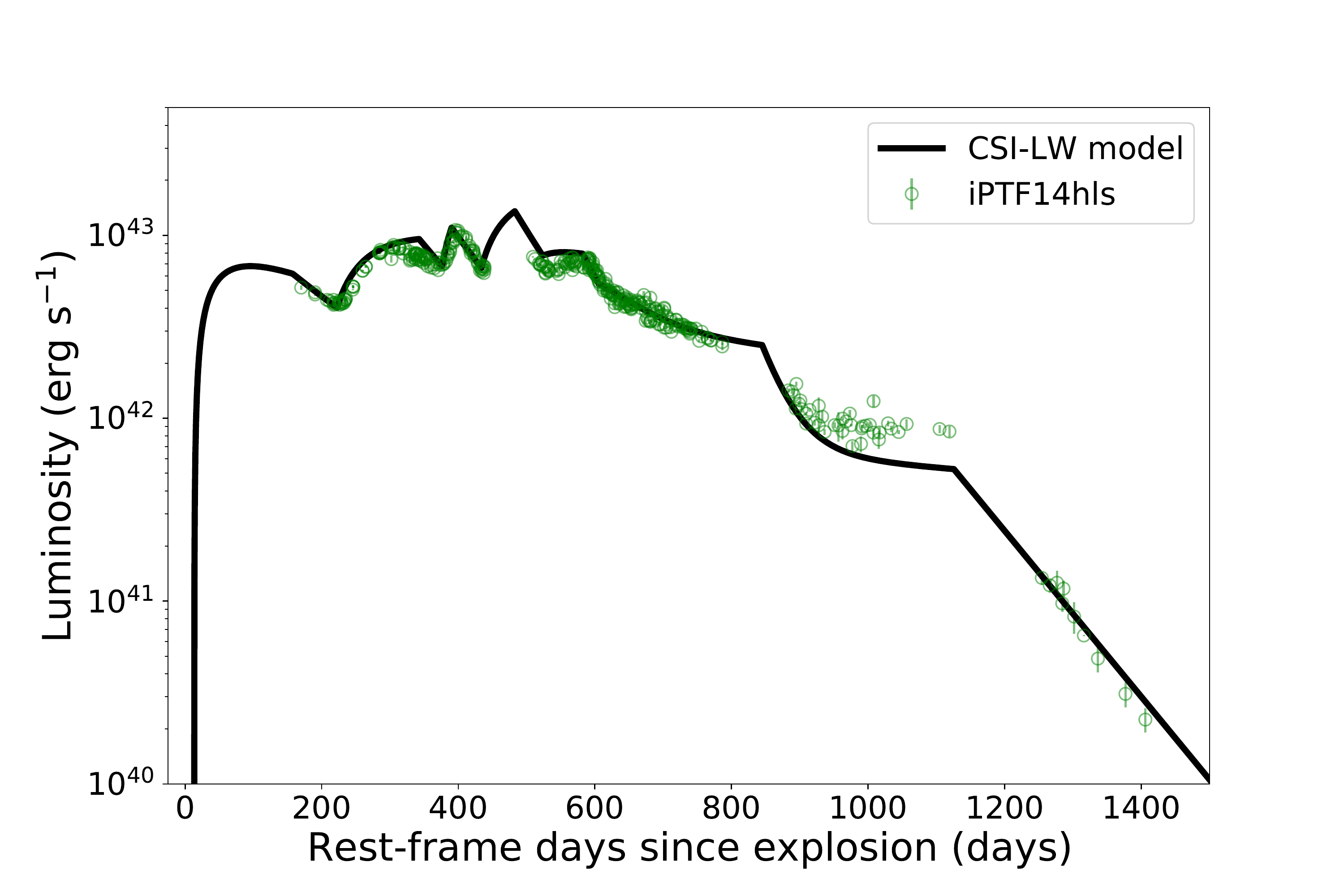} %
\includegraphics[width=1\textwidth,angle=0,width=8.9cm]{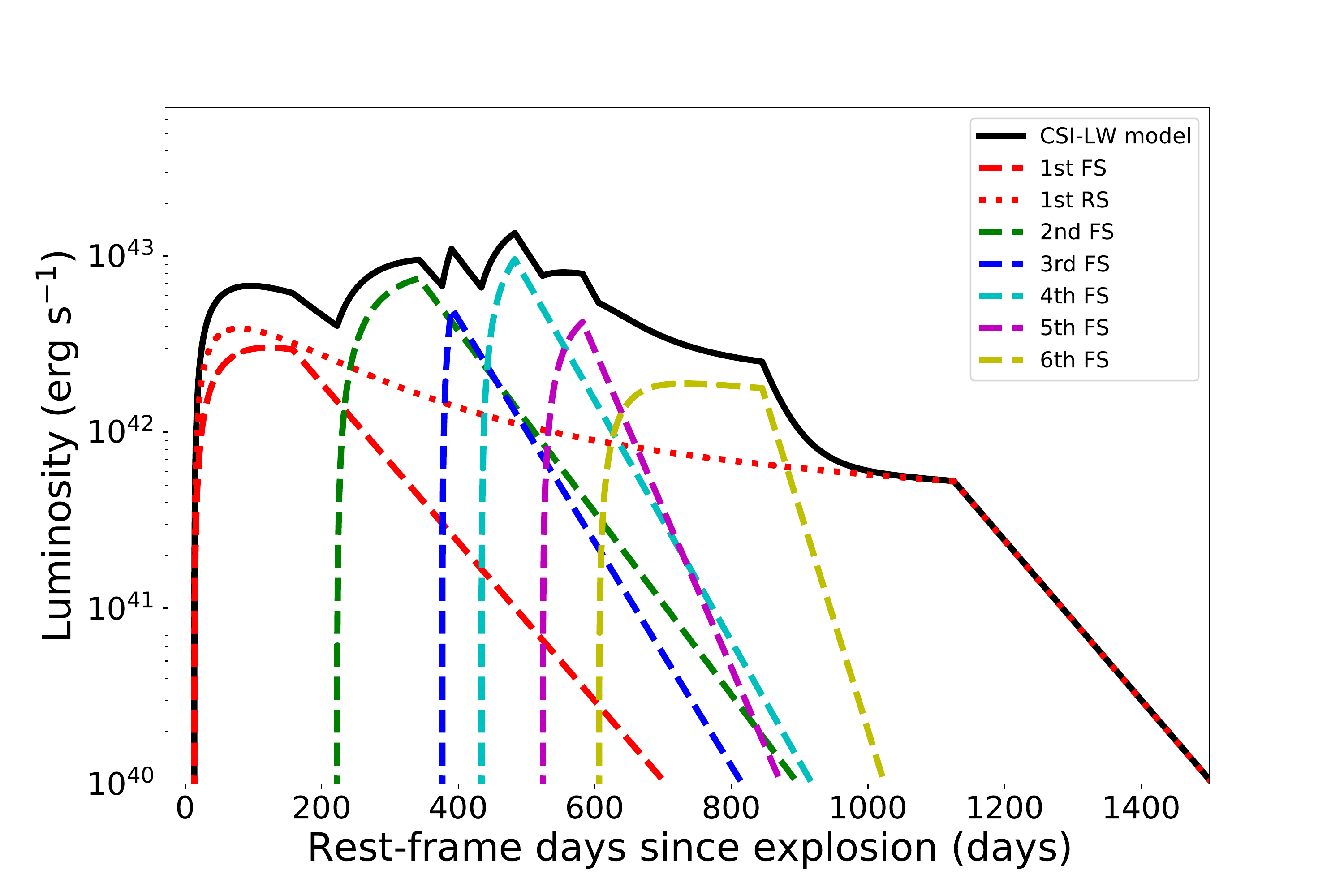} 
\centering
\caption{Fit to the bolometric luminosity (left panel) of iPTF14hls by the
CSI-LW model. The contributions from forward shocks and reverse shocks of $i$%
th CSM are plotted in right panel. FS emission is plotted in dashed lines,
while the first RS is plotted in the red dotted line. }
\label{fig:lc-in-out}
\end{figure*}

\begin{table*}[tbph]
\caption{Model parameters with the CSI plus $^{56}$Ni model.}
\label{tbl:para-in-out}
\begin{center}
\begin{tabular}{lllllll}
\hline\hline
ejecta properties &  &  &  &  &  &  \\ \hline
$M_{\mathrm{ej}} (M_{\odot})$ & 86 &  &  &  &  &  \\ 
$M_{\mathrm{Ni}}$($M_{\odot}$) & 0.9 &  &  &  &  &  \\ 
$v_{\mathrm{SN}}$ $\left( \unit{km} \unit{s}^{- 1} \right)$ & 5200 &  &  & 
&  &  \\ 
$\delta$ & 0 &  &  &  &  &  \\ 
$n$ & 7 &  &  &  &  &  \\ \hline
CSM properties &  &  &  &  &  &  \\ 
& 1st & 2nd & 3rd & 4th & 5th & 6th \\ \hline
$M_{\mathrm{CSM}} (M_{\odot})$ & 18 & 15 & 25 & 20 & 35 & 80 \\ 
$R_{\mathrm{in}} (10^{15} \unit{cm})$ & 8.4 & 14.8 & 18 & 21 & 23 & 26 \\ 
$\rho_{\mathrm{in}} \left( 10^{- 15} \unit{g } \unit{cm}^{-3} \right)$ & 8.5
& 7.3 & 5.4 & 5.2 & 4.1 & 1.1 \\ 
$\epsilon_{\mathrm{FS}}$ & 0.07 & 0.14 & 0.1 & 0.08 & 0.04 & 0.018 \\ 
$\epsilon_{\mathrm{RS}}$ & 0.26 & 0 & 0 & 0 & 0 & 0 \\ \hline
\end{tabular}%
\end{center}
\end{table*}

\subsection{CSI plus $^{56}$Ni model}

From Table \ref{tbl:para-L-W} we see that in the pure CSI model the
progenitor has a zero-age main-sequence mass $M_{\mathrm{zams}}\simeq M_{%
\mathrm{ej}}+\Sigma M_{\mathrm{CSMs}}\simeq 245.5M_{\odot }$. The true value
of $M_{\mathrm{zams}}$ may be higher than the value given here, to account
for other pre-explosion mass-loss processes and the massive compact remnant,
if the progenitor of iPTF14hls was not completely disrupted. For such
massive stars, it is expected that some amount of $^{56}$Ni is synthesized
during the SN explosion \citep{Heger02, Umeda02}.

In addition, in pure-CSI model the initial slow decline of the light curve
can only be fitted with a CSM interaction in which the RS emission dominates
over FS emission. This results in a $\epsilon _{\mathrm{FS}}=0.02$
exceptionally small compared to $\epsilon _{\mathrm{FS}}$ of other CSMs. The
initial decline can also be fitted with a CSM whose mass $M_{\mathrm{CSM}%
}\gtrsim 20M_{\odot }$, so that the FS emission can last for more time to
have a similar time dependence as RS. This is because the heating rates of
FS and RS have the same power-law index (see Equation \ref{eq:L_R-L_F}).
However, with such a massive CSM, it will be difficult to fit the
exponential decline starting on $\sim 1100$ days post explosion (see Figure %
\ref{fig:lc-in-out}) because the large total mass ($M_{\mathrm{ej}}+\Sigma
M_{\mathrm{CSMs}}$) results in a shallow exponential decline.

\begin{figure*}[tbph]
\includegraphics[width=1\textwidth,angle=0,width=8.9cm]{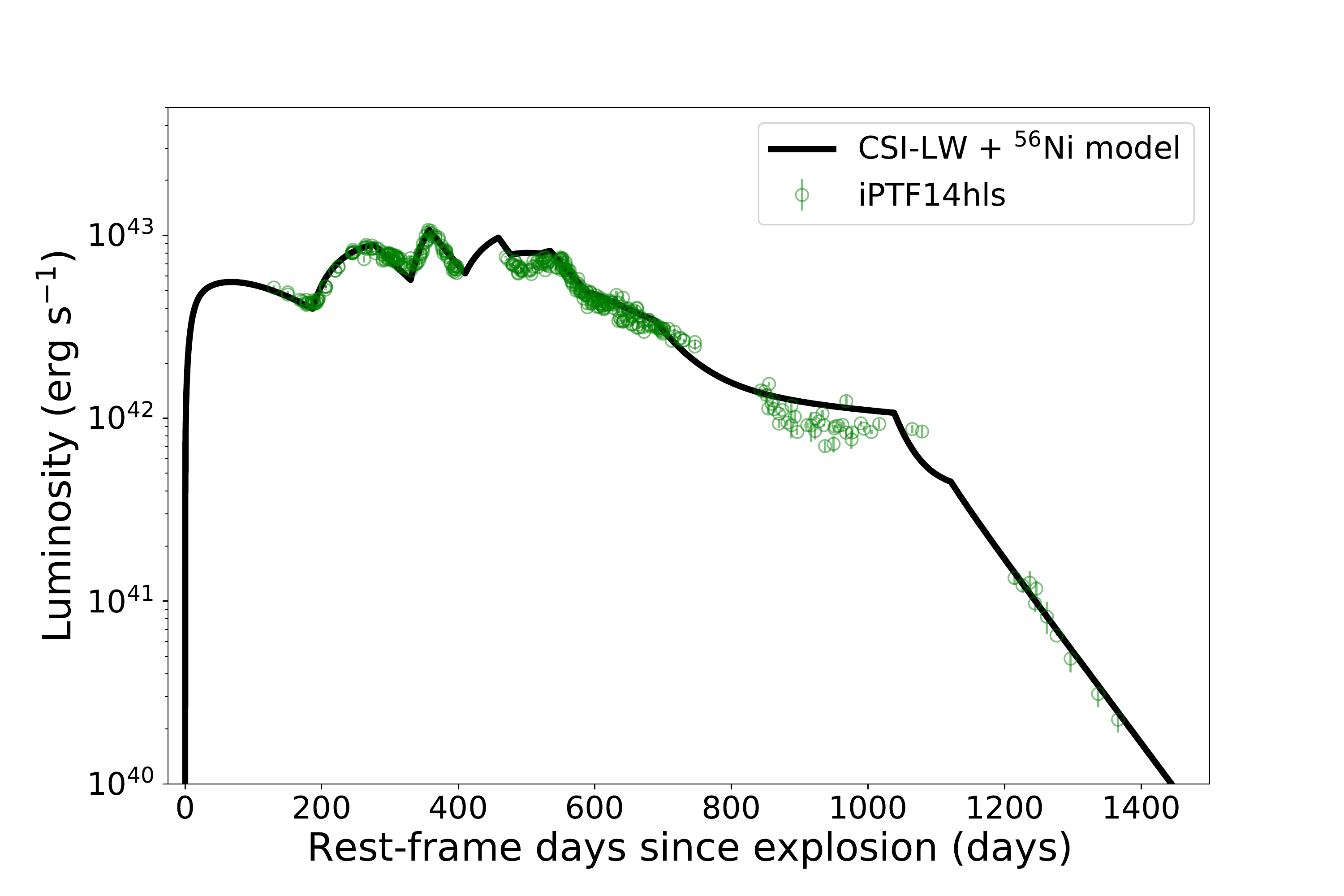} %
\includegraphics[width=1%
\textwidth,angle=0,width=8.9cm]{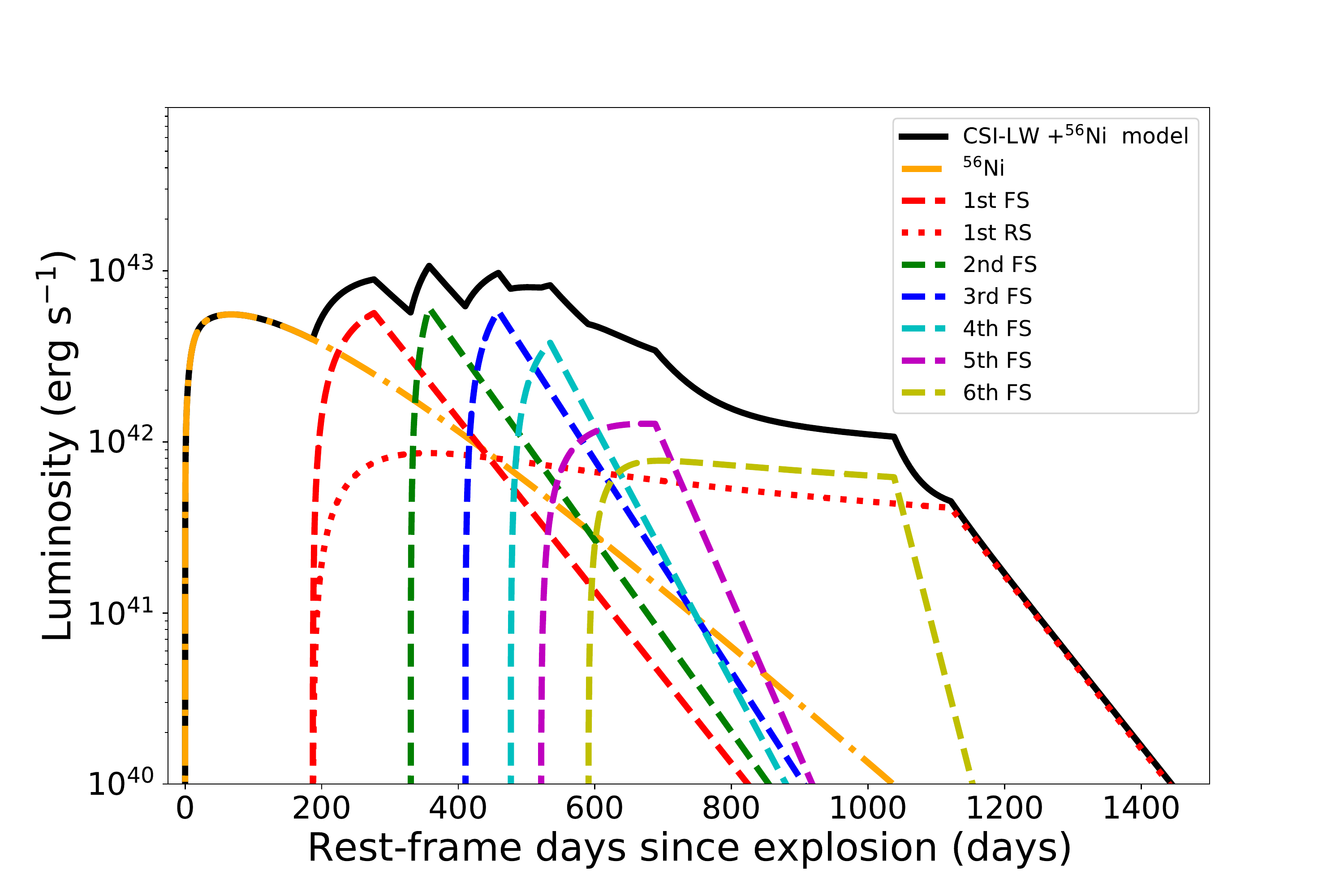} \centering
\caption{Fit to the bolometric luminosity (left panel) of iPTF14hls by the
CSI-LW plus $^{56}$Ni model. The contributions from FS, RS, and $^{56}$Ni
are plotted in right panel. FS emission are plotted in dashed lines, the
first RS is plotted in the red dotted line, and $^{56}$Ni contribution is
plotted in the orange dashed-dotted line. }
\label{fig:CSI-Ni}
\end{figure*}

Based on above considerations, we explore the possibility of fitting the
initial decline of the light curve by $^{56}$Ni decay. The fitting result is
presented in Figure \ref{fig:CSI-Ni}, with the fitting parameters listed in
the Table \ref{tbl:para-in-out}. The gray opacity to $^{56}$Ni cascade decay
photons is $\kappa _{\gamma }=0.027\unit{cm}^{2}\unit{g}^{-1}$ %
\citep{Colgate80, Swartz95}. In this model, the explosion energy is very
close to the value given in Equation $\left( \ref{eq:exp-energy}\right) $.

It is found that the observed light curve around $\sim 1000$ days post
explosion declines according to $L\propto t^{-0.3}$ (see Figure \ref%
{fig:CSI-Ni}), while the theoretical light curve with $n=7$ and $s=2$ gives $%
L\propto t^{-0.6}$. This is the reason why there is some deviation between
the theoretical light curve and observational data. Actually, $L\propto
t^{-0.3}$ is consistent with $n=12$ and $s=2$. Such an outer density profile
of the ejecta is expected for the envelopes of red supergiant progenitors,
while $n=7$ can be found in the envelopes of more compact progenitors %
\citep{Matzner99, Chatzopoulos12}. This indicates that the outer density
profile of the ejecta changes during the successive collisions with CSMs.

\section{Mass loss history}

\label{sec:mass-loss}

Hereafter we consider the CSI-LW and CSI-LW plus $^{56}$Ni models as our
fiducial models. Given the fitting parameters listed in Table \ref%
{tbl:para-L-W} and Table \ref{tbl:para-in-out}, mass loss history of
iPTF14hls can be studied. Before the discovery of iPTF14hls on 2014 Sep. 22,
there is a historic eruption at the same position of iPTF14hls on 1954 Feb.
23 \citep{Arcavi17}. Assuming that this eruption corresponds to the first
CSM listed in Table \ref{tbl:para-in-out}, the wind velocity $v_{\mathrm{wind%
}}$ can be derived, namely $v_{\mathrm{wind}}=R_{\mathrm{out},1}\left(
1+z\right) /60.58\unit{yr}\simeq 42\unit{km}\unit{s}^{-1}$, slightly higher
than the observed wind velocity of Type II-P SNe \citep{Smith14}. Here $R_{%
\mathrm{out}}$ is the outer radius of the CSM that is calculated according
to $R_{\mathrm{in}}$, $M_{\mathrm{CSM}}$ and $\rho _{\mathrm{in}}$. If the
initial decline of the light curve is attributed to $^{56}$Ni decay, the
wind velocity is $v_{\mathrm{wind}}\simeq 71\unit{km}\unit{s}^{-1}$. Based
on these wind velocities, we list the mass loss history in Table \ref%
{tbl:mass-loss}.

\begin{table*}[tbph]
\caption{Mass loss history of iPTF14hls derived according to the CSI-LW and
CSI-LW+$^{56}$Ni models.}
\label{tbl:mass-loss}
\begin{center}
\begin{tabular}{llccccccccccccc}
\hline
&  & \multicolumn{6}{l}{CSI-LW model} &  & \multicolumn{6}{l}{CSI-LW plus $%
^{56}$Ni model} \\ \hline
\multicolumn{1}{c}{} &  & 1st & 2nd & 3rd & 4th & 5th & 6th &  & 1st & 2nd & 
3rd & 4th & 5th & 6th \\ \hline
\multicolumn{1}{c}{$\dot{M}$} & $\left( M_{\odot }\unit{yr}^{-1}\right) $ & 
0.37 & 0.6 & 1.3 & 1.3 & 1.2 & 1.0 &  & 0.85 & 2.3 & 2.5 & 3.3 & 3.1 & 1.0
\\ 
\multicolumn{1}{c}{$T_{\mathrm{start}}$} & $\left( \unit{yr}\right) $ & 58.6
& 117.1 & 126.2 & 152.3 & 176.0 & 244.7 &  & 58.6 & 72.5 & 90.2 & 99.7 & 
113.9 & 193.8 \\ 
\multicolumn{1}{c}{$T_{\mathrm{end}}$} & $\left( \unit{yr}\right) $ & 4.5 & 
75.0 & 120.0 & 133.5 & 157.5 & 176.3 &  & 37.4 & 65.9 & 80.2 & 93.5 & 102.4
& 115.8 \\ 
\multicolumn{1}{c}{$T_{\mathrm{duration}} $} & $\left( \unit{yr}\right) $ & 
54.1 & 42.1 & 6.2 & 18.7 & 18.5 & 68.4 &  & 21.2 & 6.6 & 10.1 & 6.2 & 11.5 & 
78.0 \\ 
\multicolumn{1}{c}{$T _{\mathrm{waiting}}$} & $\left( \unit{yr}\right) $ & 
4.5 & 16.5 & 3.0 & 7.3 & 5.3 & 0.2 &  & 37.4 & 7.4 & 7.6 & 3.3 & 2.8 & 1.9
\\ 
\multicolumn{1}{c}{$E_{\mathrm{burst}}$} & $\left( 10^{47}\unit{erg}\right) $
& 3.6 & 4.5 & 1.4 & 4.4 & 3.9 & 12.4 &  & 9.1 & 7.5 & 12.6 & 10.1 & 17.6 & 
40.3 \\ \hline
\end{tabular}%
\end{center}
\par
\textbf{Notes.} $\dot{M}$ is the mass loss rate, $T_{\mathrm{start}}$, $T_%
\mathrm{end}$, $T_{\mathrm{duration}} $, $T _{\mathrm{waiting}}$, and $E_{%
\mathrm{burst}}$ are the start time, end time, duration, waiting time, and
burst energy of the mass loss episodes, respectively. All times are in rest
frame of the host galaxy.
\end{table*}

All timescales in Table \ref{tbl:mass-loss} are calculated in the rest frame
of the host galaxy of iPTF14hls with time zero point set on the discovery
date of iPTF14hls\footnote{%
A more appropriate zero point time may be the explosion time of iPTF14hls.
However, given the large uncertainty of the explosion time, here we just use
the discovery date. This approximation only introduces a very minor error in
the derived quantities.}. Mass loss rates are $\dot{M}\simeq 0.4-1.3M_{\odot
}\unit{yr}^{-1}$ (in the CSI-LW model) or $\dot{M}\simeq 0.85-3.3M_{\odot }%
\unit{yr}^{-1}$ (in the CSI-LW plus $^{56}$Ni model), slightly higher than
that of Type IIn \citep{Schlegel90, Filippenko97} and IIn-P SNe 
\citep[e.g.,][]{Kiewe12,
Taddia13, Smith14}.

In Table \ref{tbl:mass-loss} we also list the outburst energy for each
mass-loss episode, calculated according to $E_{\mathrm{burst}}=\frac{1}{2}M_{%
\mathrm{CSM}}v_{\mathrm{wind}}^{2}$. The burst energies have a typical value
of $10^{47}-10^{48}\unit{erg}$. These derived burst energies are small
compared to the gravitational binding energies (see Equation \ref%
{eq:E-binding}\ below) of the CSM-shells before ejection. This indicates
that there could be much more energy-release episodes during the final
evolution of the progenitor, of which only a fraction of the energy-release
episodes trigger large-scale mass loss.

\begin{figure*}[tbph]
\includegraphics[width=1\textwidth,angle=0,width=8.9cm]{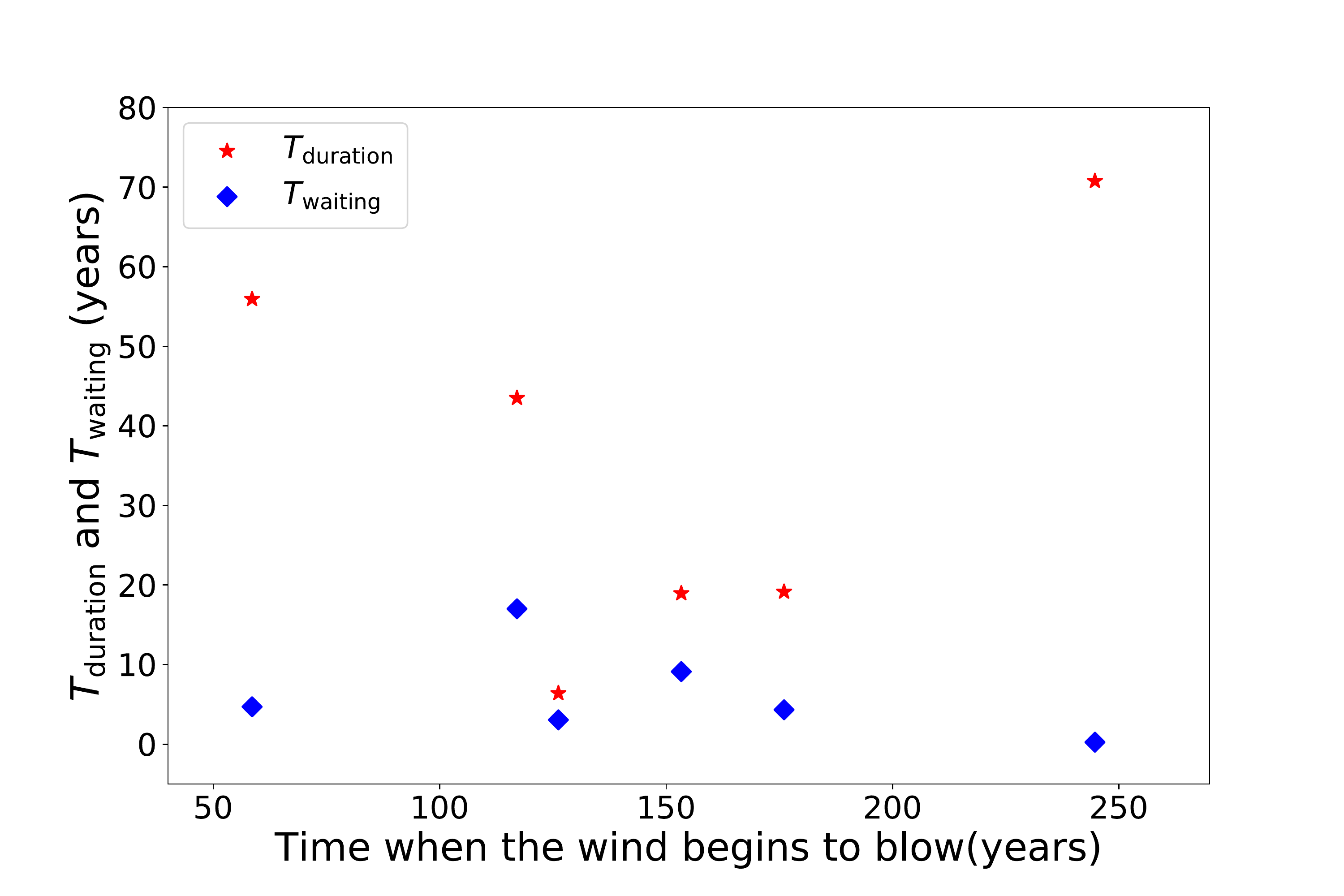} %
\includegraphics[width=1\textwidth,angle=0,width=8.9cm]{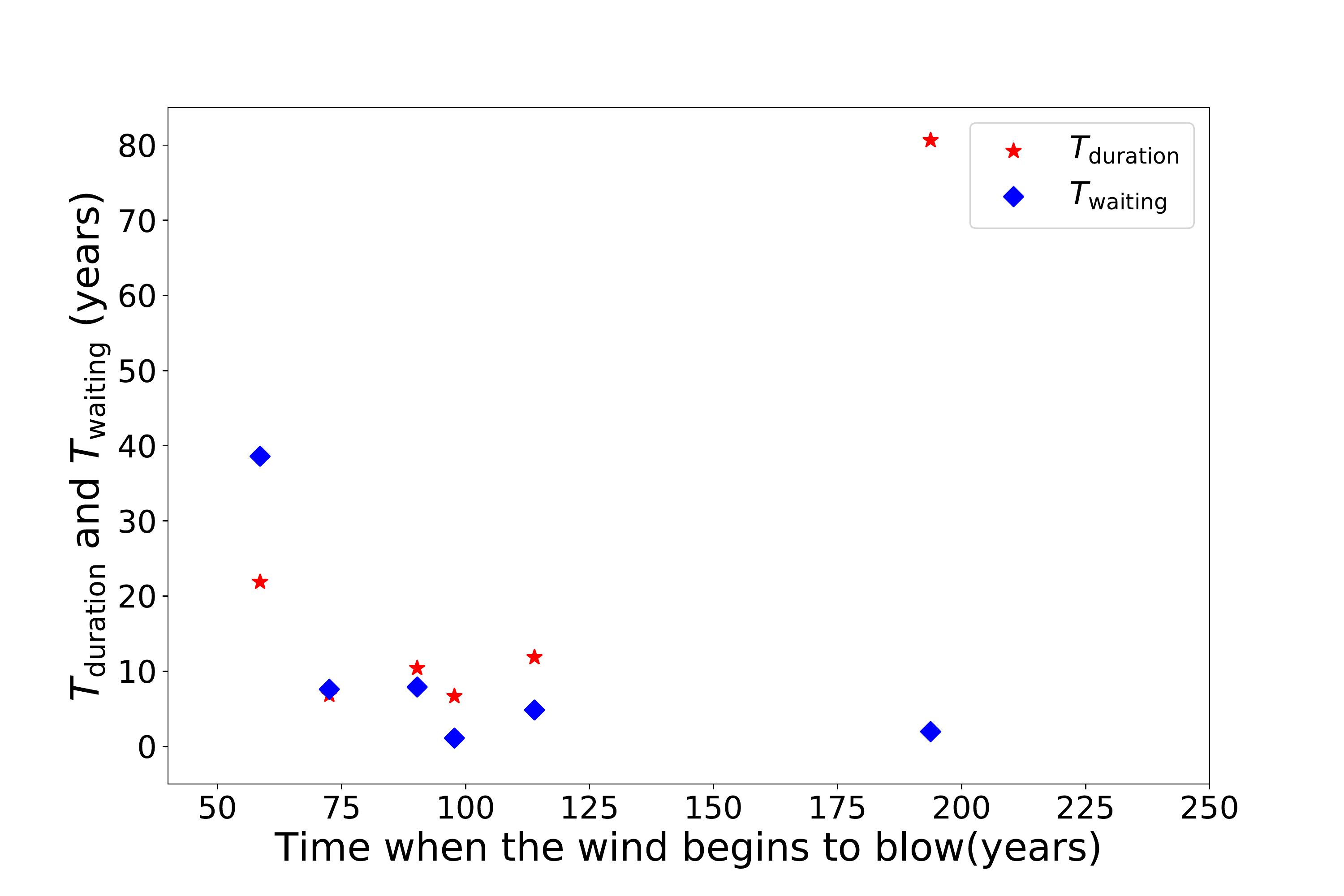}
\caption{The waiting time and duration of the mass-loss episodes in the
CSI-LW (left panel) and CSI-LW+$^{56}$Ni (right panel) models.}
\label{fig:waiting}
\end{figure*}

Waiting times are calculated as $T_{\mathrm{{waiting},}i}=T_{\mathrm{{end},}%
i}-T_{\mathrm{{start},}i-1}$ for the $i$th CSM. The duration and waiting
time of the mass-loss episodes are presented in Figure \ref{fig:waiting}.
From Figure \ref{fig:waiting}, we see that during the final 200 years of
evolution of the progenitor of iPTF14hls, mass loss is very frequent, with
only very short quiescence periods (waiting times) following the longer wind
blowing episodes. An interesting exception is the last mass-loss episode
(corresponding to the first CSM) in the CSI-LW+$^{56}$Ni model, where the
waiting time $T_{\mathrm{waiting}}=37.4\unit{yr}$ is longer than the
duration $T_{\mathrm{duration}}=21.2\unit{yr}$. This waiting time is also
much longer than the waiting times of earlier mass-loss episodes, which have
values $T_{\mathrm{waiting}}=1.9-7.6\unit{yr}$ (see Figure \ref{fig:waiting}
and Table \ref{tbl:mass-loss}). If the CSI-LW+$^{56}$Ni model is correct,
this indicates that there could be a long quiescence before the final
collapse of the progenitor.

It should be mentioned that the wind velocity is expected to increase as the
progenitor becomes more and more compact via shedding a large fraction of
its mass. Taking this fact into account, the mass loss rates of the earlier
mass-loss episodes (the second to sixth CSMs listed in Table \ref%
{tbl:mass-loss}) could be lower and the duration and waiting time could be
longer than the values given in Table \ref{tbl:mass-loss}. It is also
possible that one or more of the CSM-shells are the merging results with the
late fast-moving shell running into the earlier slow-moving shell.

\section{Discussion and conclusions}

\label{sec:conc}

The derived main-sequence mass ($M_{\mathrm{zams}}\simeq 245.5M_{\odot }$)
is consistent with a pair-instability supernova 
\citep[PISN;][]{Barkat67, Rakavy67, Ober83, Bond84,
Glatzel85, Scannapieco05, Kasen11}, which has a mass range $140M_{\odot }<M_{%
\mathrm{zams}}<260M_{\odot }$. However, theoretical studies suggest that
massive stars with $M_{\mathrm{zams}}\simeq 245.5M_{\odot }$ will end their
lives during the final explosions that disrupt themselves completely and do
not episodically lose mass. In this case the CSM-shells listed in Table \ref%
{tbl:para-in-out} may be generated by envelope instabilities 
\citep{Yoon10,
Owocki15}, which is suggested for the mass loss of some Type II-P SNe in
their final evolution stage \citep[e.g.,][]{Rui19}. Another difficulty for
the interpretation of iPTF14hls as a PISN is that the derived $^{56}$Ni mass 
$M_{\mathrm{Ni}}=0.9M_{\odot }$ (see Table \ref{tbl:para-in-out}) in the
CSI-LW+$^{56}$Ni model is lower than the expected yield of $^{56}$Ni during
the final disruptive explosion of a PISN, which could have $M_{\mathrm{Ni}%
}\sim 50M_{\odot }$ for a $M_{\mathrm{zams}}\simeq 250M_{\odot }$ star %
\citep{Heger02, Umeda02}.

The properties of the progenitor of iPTF14hls are more consistent with a
pulsational pair-instability supernova\footnote{%
Because of mass loss, the metallicity of the progenitor of iPTF14hls should
be $Z\leq 0.5Z_{\odot }$ 
\citep{Langer07, Georgy13, Yusof13, Spera15,
Leung19} to form He core massive enough to undergo pulsational
pair-instability. This results in an opacity $\kappa $ essentially identical
to our adopted value because electron scattering opacity depends on
metallicity very weakly.} 
\citep[PPISN;][]{Barkat67, Woosley07, Chen14,
Woosley15}, as discussed in the literature \citep{Arcavi17, Woosley18}.
There may be some overlap between the initial mass of a star that will end
up as a PISN or PPISN, which has a mass range $70M_{\odot }<M_{\mathrm{zams}%
}<140M_{\odot }$. This is plausible because our current understanding of
massive star evolution is incomplete owing to the complex physics involved,
including metallicity-related mass loss, chemical mixing, rotation,
binary-interaction-related mass transfer, and magnetic torques 
\citep{Ohkubo09, ChatzopoulosWheeler12, Woosley17, Woosley19, Leung19,
Marchant19}. Indeed, \cite{Yoshida16} calculated the evolution of stars with
metallicity $Z=0.004$ and initial masses of 140, 200, and $250M_{\odot }$,
which are well within the usually assumed mass range of PISNe. They found
their masses decrease to 54, 59, and $61M_{\odot }$ before the neon burning
owing to mass loss. Eventually, the above three stars experience six ($%
M=140M_{\odot }$) to three ($M=250M_{\odot }$) episodes of mass loss induced
by PPI during their final $1$ ($M=140M_{\odot }$) to $1400$ $(M=250M_{\odot
})$ $\unit{yr}$ evolution before core collapse.

For a star that will end up as a PPISN, electron--positron pairs will be
generated when the central temperature of the massive helium core rises to $%
(1.8-2.3)\times 10^{9}\unit{K}$ \citep{Fowler64, Barkat67, Rakavy67}. Sudden
loss of pressure due to the production of electron--positron pairs leads to
the contraction of the helium core. Then explosive burning of oxygen ensues
so as to eventually reverse the contraction of the core to expansion. Such
pulsational activity becomes more and more energetic to shed some amount of
envelope during the final evolution of a PPISN progenitor. The duration of
pulsational activity spans a wide range of time, from a few hours to $10^{4}%
\unit{yr}$ \citep{Woosley17}, in accord with the mass loss history listed in
Table \ref{tbl:mass-loss}.

During the pair pulsation, the progenitor of a PPISN may be a red supergiant
(RSG), a blue supergiant (BSG), a luminous blue variable, or Wolf--Rayet
star if the hydrogen envelope is completely shed away by continuous mass
loss \citep{Woosley17}. The gravitational binding energy of the hydrogen
shell that will be blown away by the pulsation is%
\begin{equation}
E_{p}=1.2\times 10^{49}\unit{erg}\left( \frac{M_{R}}{150M_{\odot }}\right)
\left( \frac{M_{s}}{30M_{\odot }}\right) \left( \frac{R_{p}}{10^{14}\unit{cm}%
}\right) ^{-1},  \label{eq:E-binding}
\end{equation}%
where $M_{s}$ is the shell mass that will be blown away, $M_{R}$ is the mass
left over after mass shedding, $R_{p}$ is the radius of the progenitor star
during helium burning. $R_{p}\approx 10^{14}\unit{cm}$ if the progenitor
star is a RSG, while $R_{p}\approx 10^{13}\unit{cm}$ for a BSG. The
progenitor of iPTF14hls is likely an RSG since the above binding energy is
closer to the burst energies listed in Table \ref{tbl:mass-loss}. Owing to
the same reason, the initial decline of the light curve could be more likely
attributed to $^{56}$Ni decay since in the CSI-LW+$^{56}$Ni model the burst
energies is closer to the binding energy.

The amount of $^{16}$O burnt during one pulsation can be estimated as%
\begin{equation}
M_{\mathrm{O}}=0.01M_{\odot }\left( \frac{E_{p}}{10^{49}\unit{erg}}\right) .
\end{equation}%
Here the energy release by $^{16}$O burning is $0.52\unit{MeV}$ per nucleon %
\citep{Arnett96, Wallerstein97}. The contraction timescale of the helium
core is the dynamical timescale. Therefore such an amount of $^{16}$O will
be burnt in%
\begin{equation}
t_{d}=3.8\unit{hr}\left( \frac{R_{c}}{10^{12}\unit{cm}}\right) ^{3/2}\left( 
\frac{M_{c}}{50M_{\odot }}\right) ^{-1/2},  \label{eq:dyn-time}
\end{equation}%
where $R_{c}$ and $M_{c}$ are the radius and mass of the helium core that
participates in the pulsation. The timescales of mass ejection and
subsequent resettlement to a new bound state are the Kelvin-Helmholtz
timescale, i.e., several years (see Table \ref{tbl:mass-loss}), which is
much longer than the dynamical timescale given by Equation $\left( \ref%
{eq:dyn-time}\right) $.

It is usually assumed that the remnants of PPISNe are massive black holes
and no $^{56}$Ni is ejected to power the SN light curve %
\citep[e.g.,][]{Woosley17, Leung19}. \cite{Tolstov17}, on the other hand,
propose that a PPISN can eject a large amount of $^{56}$Ni ($M_{\mathrm{Ni}%
}=6M_{\odot }$) to power the light curve of SLSN PTF12dam, despite the
formation of a black hole in PTF12dam. Given the moderate amount of $^{56}$%
Ni in the CSI-LW+$^{56}$Ni model (see Table \ref{tbl:para-in-out}), we
suggest that the $^{56}$Ni may be ejected during the collapse of the helium
core to a massive black hole, i.e., a collapsar \citep{Woosley93}. This
implies a rapid rotation of the helium core. However, such a rapid rotation
is in tension with the existence of the hydrogen-rich envelope of the
progenitor of iPTF14hls since rotation tends to remove the hydrogen
envelope. This tension can be relaxed by assuming a differential rotation of
the progenitor star. The derived explosion energy (Equation \ref%
{eq:exp-energy}) is also consistent with the collapsar model.

To date, there are several candidates for PPISNe 
\citep[][]{BenAmi14, Lunnan18,
Gomez19}. \cite{Nicholl20} reported the detection of a Type IIn SLSN 2016aps
with energy $E_{\mathrm{SN}}\gtrsim 10^{52}\unit{erg}$ and a total mass
(ejecta\thinspace +\thinspace CSM) exceeding $50-100M_{\odot }$. Detailed
one-dimensional radiation-hydrodynamic simulations \citep{Suzuki21} suggest
that SN 2016aps could be a collision of an ejecta mass $M_{\mathrm{ej}%
}=30M_{\odot }$ with a $M_{\mathrm{CSM}}\simeq 8M_{\odot }$ wind-like CSM of
outer radius $10^{16}\unit{cm}$. SN 2016aps has a radiated energy $E_{%
\mathrm{rad}}\gtrsim 5\times 10^{51}\unit{erg}$, which is an order of
magnitude higher than iPTF14hls. The mass loss rate of SN 2016aps, $\dot{M}%
\approx 0.3M_{\odot }\unit{yr}^{-1}$, assuming a wind velocity $u_{w}=100%
\unit{km}\unit{s}^{-1}$, is comparable to that of iPTF14hls. \cite{Yang21}
reported the Type II SN 2020faa, whose first six months of light curves are
of great similarity with those of iPTF14hls.

The light curve of iPTF14hls is very unusual among SNe studied so far. In
this paper, we explore the possibility of interpreting iPTF14hls as a
multiple interaction-powered SN. We find that within reasonable parameters,
the theoretical light curve matches well with the light curve of iPTF4hls.
This makes iPTF14hls a possible candidate for PPISN. \cite{Moriya20}
proposed that iPTF14hls is produced by a continuous outflow like a stellar
wind rather than a mass ejection. They calculated the mass-loss rates of
iPTF14hls as high as $10M_{\odot }\unit{yr}^{-1}$ in the bright phase. Such
an extreme mass loss rate is much higher than the results obtained from our
calculations. \cite{Dessart18} found magnetar-powered SN ejecta reproduces
some the observed properties of iPTF14hls, including the sustained
brightness in the $R$ band, the blue optical color, and the broad HI lines.
However, the magnetar model is difficult to produce fluctuating light curves
with multiple peaks, unless consider variable thermal energy injection from
magnetar spin down \citep{Moriya22}. In the future, more observations and
improved theoretical modelings can improve our understanding of very massive
stars \citep{Heger03}.

\begin{acknowledgements}
We thank the anonymous referee for his/her constructive
comments that allow us to improve the paper significantly.
We also thank Iair Arcavi and Jesper Sollerman for providing us the observational data.
This work is supported by the National Program on Key Research and Development
Project of China (Grant Nos. 2021YFA0718500 and 2017YFA0402600), and the National Natural Science Foundation of China (Grant
Nos. U1938201, 11833003, 12103055). X.F. Wang is supported by the National Natural Science 
Foundation of China (NSFC grants 12033003 and 11633002), 
the Major State Basic Research Development Program (grant 2016YFA0400803), 
and the Scholar Program of Beijing Academy of Science and Technology (DZ:BS202002).
\end{acknowledgements}

\end{document}